\documentclass[10pt]{bmc_article}

\usepackage{cite} % Make references as [1-4], not [1,2,3,4]
\usepackage{url}  % Formatting web addresses
\usepackage{ifthen}  % Conditional
\usepackage{multicol}   %Columns
\usepackage[utf8]{inputenc} %unicode support
\usepackage{amssymb,latexsym,amsmath,amscd,amsthm}

\usepackage{subfig}% per fer subfloat a les figures (esther)
\usepackage{graphicx}% per incloure els grafics
\usepackage{color} %per escriure en colors (util per corregir)
\newboolean{publ}

\newenvironment{bmcformat}{\baselineskip20pt\sloppy\setboolean{publ}{false}}{\baselineskip20pt\sloppy}

\begin{document}
\begin{bmcformat}


\title{EM for phylogenetic topology reconstruction on nonhomogeneous data}

%%%%%%%%%%%%%%%%%%%%%%%%%%%%%%%%%%%%%%%%%%%%%%
%%                                          %%
%% Enter the authors here                   %%
%%                                          %%
%% Ensure \and is entered between all but   %%
%% the last two authors. This will be       %%
%% replaced by a comma in the final article %%
%%                                          %%
%% Ensure there are no trailing spaces at   %%
%% the ends of the lines                    %%     	
%%                                          %%
%%%%%%%%%%%%%%%%%%%%%%%%%%%%%%%%%%%%%%%%%%%%%%

\author{Esther Ib\'{a}\~{n}ez-Marcelo\correspondingauthor$^1$%
         \email{Esther Iba\~{n}ez\correspondingauthor - eibanez@crm.cat}
       and
         Marta Casanellas$^2$%
         \email{Marta Casanellas - marta.casanellas@upc.edu}%
      }

%%%%%%%%%%%%%%%%%%%%%%%%%%%%%%%%%%%%%%%%%%%%%%
%%                                          %%
%% Enter the authors' addresses here        %%
%%                                          %%
%%%%%%%%%%%%%%%%%%%%%%%%%%%%%%%%%%%%%%%%%%%%%%

\address{%
    \iid(1)Centre de Recerca Matem\'atica, Campus de Bellaterra, Edifici C - 08193 Bellaterra (Barcelona), Spain\\
    \iid(2)Departament Matem\'atica Aplicada I, Universitat Polit\`ecnica de Catalunya, %
        Avda. Diagonal 647, Barcelona, Spain
}%

\maketitle

%%%%%%%%%%%%%%%%%%%%%%%%%%%%%%%%%%%%%%%%%%%%%%
%%                                          %%
%% The Abstract begins here                 %%
%%                                          %%
%% Please refer to the Instructions for     %%
%% authors on http://www.biomedcentral.com  %%
%% and include the section headings         %%
%% accordingly for your article type.       %%
%%                                          %%
%%%%%%%%%%%%%%%%%%%%%%%%%%%%%%%%%%%%%%%%%%%%%%

\begin{abstract}
        % Do not use inserted blank lines (ie \\) until main body of text.

Background:
The reconstruction of the phylogenetic tree topology of four taxa is, still nowadays, one of the main challenges in phylogenetics. Its difficulties lie in considering not too restrictive evolutionary models, and correctly dealing with the long-branch attraction problem. The correct reconstruction of 4-taxon trees is crucial for making quartet-based methods work and being able to recover large phylogenies.

Results:
In this paper we consider an expectation-maximization method for maximizing the likelihood of (time nonhomogeneous) evolutionary Markov models on trees. We study its success on reconstructing 4-taxon topologies and its performance as input method in quartet-based phylogenetic reconstruction methods such as QFIT and QuartetSuite. Our results show that the method proposed here outperforms neighbor-joining and the usual (time-homogeneous continuous-time) maximum likelihood methods on 4-leaved trees with among-lineage instantaneous rate heterogeneity, and perform similarly to usual continuous-time maximum-likelihood when data satisfies the assumptions of both methods.

Conclusions:
The method presented in this paper is well suited for reconstructing the topology of any number of taxa via quartet-based methods and  is highly accurate, specially regarding largely divergent trees and time nonhomogeneous data.

\end{abstract}

\ifthenelse{\boolean{publ}}{\begin{multicols}{2}}{}

%%%%%%%%%%%%%%%%%%%%%%%%%%%%%%%%%%%%%%%%%%%%%%
%%                                          %%
%% The Main Body begins here                %%
%%                                          %%
%% Please refer to the instructions for     %%
%% authors on:                              %%
%% http://www.biomedcentral.com/info/authors%%
%% and include the section headings         %%
%% accordingly for your article type.       %%
%%                                          %%
%% See the Results and Discussion section   %%
%% for details on how to create sub-sections%%
%%                                          %%
%% use \cite{...} to cite references        %%
%%  \cite{koon} and                         %%
%%  \cite{oreg,khar,zvai,xjon,schn,pond}    %%
%%  \nocite{smith,marg,hunn,advi,koha,mouse}%%
%%                                          %%
%%%%%%%%%%%%%%%%%%%%%%%%%%%%%%%%%%%%%%%%%%%%%%

%%%%%%%%%%%%%%%%
%% Background %%
%%
\section*{Background}
 %\cite{koon,oreg,khar,zvai,xjon,schn,pond,smith,marg,hunn,advi,koha,mouse}

Obtaining a good method for reconstructing the phylogenetic topology of four taxa is one of the crucial goals in phylogenetics. The four-taxon trees, if correctly inferred, can be used as input of quartet-based methods in order to reconstruct larger trees. But due to the complexity of real data, the problem of reconstructing four-taxon trees is not so easy. Most phylogenetic reconstruction methods assume simple evolutionary models that may not really fit real data, which leads to incorrect phylogenetic inference (\cite{Kelchner2007}, \cite{Ripplinger2010}, \cite{Jermiin2004}, \cite{Galtier1998}, \cite{YY99}). For example, many of them rely on continuous-time Markov processes with a constant instantaneous mutation rate matrix along the tree (the so-called global \textit{homogeneity}), and also assume time-reversibility (and hence stationarity). On the other hand, as pointed out in \cite{Ranwez2001}, in order to make quartet-based methods work it is extremely important to obtain 4-taxon tree 
reconstruction 
methods that are not affected by the presence of long-branch attraction \cite{Anderson2004}.

Most evolutionary models are described by a Markov process over the tree, that is, conditional rates of change at two diverging sequences depend only on the current state and are independent of the previous sates \cite[chapter 8.2]{Semple2003}. 
\textit{Markov processes} on trees are specified by a distribution at the root of the tree and a transition matrix at each branch and, in contrast to continuous-time models, the Markov process on each branch is not assumed to be time-homogeneous \cite{Jayaswal2005} (that is, they are \textit{locally} heterogeneous). These models directly consider as parameters the entries of the substitution matrices and the root distribution (see\cite[$\S$ 4.2]{AllmanRhodes_chapter4}, \cite[chapter 8]{Semple2003}).
Barry and Hartigan \cite{barryhartigan87trans} considered such a general Markov model (henceforth called GMM), which does not assume any other constraints. In particular, it is locally and \textit{globally} time nonhomogeneous (instantaneous substitution rates are not constant on any edge, nor on the whole tree) and it is neither time-reversible nor stationary. The only restriction underlying this model is that sites evolve independently and are identically distributed (i.i.d. hypothesis). Considering this model and its submodels is one way of covering more general scenarios, in contrast to those phylogenetic methods that implement time-homogeneous and time-reversible continuous-time models (GTR and its submodels) cf. \cite{Jayaswal2005}.
%In contrast, most model-based phylogenetic methods implement homogeneous and time-reversible continuous-time models (GTR and its submodels)
%discrete-time Markov processes on trees  is one way of covering more general scenarios,
%because  one  implicitly allows different instantaneous mutation rate matrices across lineages (thus avoiding the homogeneity hypothesis) and one does not necessarily assume stationarity or time-reversibility. 

%the exponential of a rate matrix . 
 
%When no further constrains  are imposed on the transition matrices or the root distribution, this most general model is known as the \textit{general Markov model} (or Barry-Hartigan model \cite{barryhartigan87trans}) and 
The GMM above  accounts for 12 parameters per edge plus three parameters for the root distribution. When some symmetries on the transition matrices or on the root distribution are imposed, one obtains the substitution matrices of the corresponding Jukes-Cantor and Kimura (2 and 3 parameters) models among others (\cite[$\S$ 4.2]{AllmanRhodes_chapter4},\cite{Evans1993}, see Methods section). For instance, the Markov version of the K81  model \cite{Kimura1981} (henceforth referred to as K81*) deals with 3 parameters per edge (one for the conditional probability of transitions, and two for the two types of transversions, see the Methods section) which makes a total of $3*(2n -3)$ parameters in unrooted trivalent trees with $n$ leaves, whereas the usual time-homogeneous continuous-time version accounts for 2 parameters for a normalized instantaneous rate matrix constant over the tree plus one parameter per edge length (that is, $2+(2n-3)$ parameters on an unrooted trivalent tree). Notice, however, that if one 
considers a time nonhomogeneous continuous-time Kimura 3-parameter model, then the number of parameters is exactly the same as for K81*.
In this case, the only difference between both models is that K81* does not even assume local homogeneity (that is, time homogeneity over each branch), while all time-continuous models do.
%substitution matrices of the latter are more general because they do not necessary have a real logarithm (whereas those of the former do), see \cite{CK}. 
The huge amount of parameters for nonhomogeneous models  makes a maximum-likelihood approach unfeasible and 
unreliable for a whole tree on $n$ taxa if $n$ is large.  %(it would be probably unreliable for $n=10$ due to inaccuracies on estimates). 
Nevertheless, there is some hope that these more general models lead to accurate methods on 4-taxon trees. In our setting, we only deal with substitutions of nucleotides (not aminoacids) on 4-taxon trees and we will always assume  the i.i.d. hypothesis (thus excluding the possibility of heterogeneity across sites). 

In this paper we develop a maximum likelihood framework for inferring the best  tree topology under 
%these discrete-time evolutionary models. 
(general) Markov processes. Our approach is based on the widely used Expectation-Maximization algorithm.
The Expectation-Maximization algorithm (EM), as introduced in \cite{Dempster1977}, is an iterative algorithm for finding maximum likelihood estimates of parameters in statistical models that deal with unobserved data. We have adapted this algorithm to the case of phylogenetic 4-taxon trees in what we call \texttt{EMtree}. EM iteratively gives an expectation of the distribution of the nucleotide sequences at the interior nodes (this is called the E-step) and finds the parameters that maximize the likelihood for these data in the so-called M-step (because the parameters for which the maximum likelihood is achieved can be computed in a closed form for complete data). The EM algorithm has been applied to many other disciplines (see for example \cite{mclachlan2007}).

The use of the EM algorithm to estimate the continuous parameters of a phylogenetic tree under a Markov process (namely, the root distribution and the entries of the transition matrices) has been already discussed in \cite{EMpar} and \cite{Jayaswal2005}. In this paper we focus on the use of \texttt{EMtree} for estimating the tree topology for four taxa and test its performance in reconstructing the correct tree topology on simulated data. We shall see that, although it is a time consuming algorithm, it can lead to very accurate results.

First of all we test it as a 4-taxon tree reconstruction method by using the tree space proposed by Huelsenbeck \cite{Huelsenbeck1995} on data simulated from a general Markov process first and then restricting to a time-homogeneous (continuous-time)  process. %a discrete-time Markov process 
We compare \texttt{EMtree} to neighbor-joining and to the usual (continuous-time) maximum likelihood approach under both global homogeneity and nonhomogeneity (note that throughout the experiments we will restrict to time-reversible models to simulate and to recover trees). 
%Implementations of nonhomogeneous continuous-time models are scarce, and here we use the nonhomogeneous model K80 of \cite{YY99}. 
As all models considered will be stationary (and with uniform stationary distribution), we will not be evaluating compositional heterogeneity but only the effect of the variation of substitution rates among lineages \cite{Ho2004}.
%(these last two methods under the homogeneity assumption). 
Afterwards, we use these three methods as input of two quartet-based methods: one weighted (Willson's QuartetSuite \cite{Willson1999}, \cite{QuartetSuite}) and one unweighted (Maximum Quartet Fit, QFIT as implemented in Clann \cite{Creevey2005}); and compare their performance on the 12-taxon trees proposed in \cite{Ranwez2001}. Specially for largely divergent trees, we observe that \texttt{EMtree} gives the best results and is less subject to long-branch attraction.

\section*{Results and discussion}

In this section we present the results of the topology reconstruction method \texttt{EMtree}  on simulated data evolving both under homogeneous and nonhomogeneous Markov processes.

%\subsection*{Results on a tree space on four-leaves}

In figure 1
% \ref{fig:comparacio}
we present the performance of three different topology reconstruction methods on the 4-taxon tree space proposed by Huelsenbeck \cite{Huelsenbeck1995} (see Methods section): figure 1(a)
% \ref{fig:comparacio}(a)
for the usual neighbor-joining (\texttt{NJ}) with the K81 distance, 1(b)
% \ref{fig:comparacio}(b)
for the usual maximum likelihood tree  (\texttt{ML}) assuming time-homogeneous continuous-time K81 model, and 1(c)
% \ref{fig:comparacio}(c)
\texttt{EMtree} for K81* model. The Huelsenbeck tree space covers a wide range of branch lengths for 4-taxon trees: parameters $a$ and $b$ denote branch lengths as indicated in figure 2 and they were varied from 0.01 to 0.75. For each pair $(a,b)$ we draw a black cell if the topology is  correctly reconstructed on the 100 simulated alignments; a white cell denotes less than 33\% success; and the gray scale used in between is shown in the figures. The alignments were generated following the K81* model (see the Methods section) and had length 300 (fig. 1
%\ref{fig:comparacio}
left) and 1000 bp (fig. 1
% \ref{fig:comparacio}
right).

The results show that \texttt{NJ} has difficulties in the Felsenstein zone where the long-branch attraction problem is present (that is, small $a$ and large $b$) and \texttt{ML} fails for largely divergent trees (note that  trees here have at most 2.25 pairwise divergence, whereas similar studies have even used divergences of about 7.0 expected substitutions per site \cite{Ho2004}).
The \texttt{EMtree} algorithm seems to overcome these two difficulties and is clearly more accurate than \texttt{NJ} and \texttt{ML} in this tree space. Nevertheless, we need to point out that the data were simulated according to K81* model, which  fits the assumptions of \texttt{EMtree} but not of \texttt{NJ} nor \texttt{ML} methods (indeed, these last two methods are  based on the continuous-time Kimura model and assume homogeneous mutation rates along lineages). This model misspecification for \texttt{ML} leads to in incorrect inference of parameters, which is even more extreme for long branches \cite{Ho2004} (this justifies the bad performance of \texttt{ML} in the lower right corner of fig. 1(b)). To confront this situation, we also generated data under time-homogeneous continuous-time K81 model. The results are depicted in the left column of figure 3 and clearly show a similar success of \texttt{ML} and \texttt{EMtree} in this case.

In order to compare \texttt{EMtree} to time nonhomogeneous but continuous-time \texttt{ML}, we generated data 
%that matched both methods. As Kimura 2-parameter substitution matrices  (henceforth, K80 model \cite{Kimura1980}) can be always written as a continuous-time model \cite[Remark 4.3]{CK}, this model matches both approaches: \texttt{EMtree} (on its K80 version) and a nonhomogeneous continuous-time K80 model described in \cite{YY99}. 
evolving under a continuous-time nonhomogeneous K81 model. This is a special case of K81* and therefore this model matches both approaches: \texttt{EMtree} (on its K81* version) and \texttt{bppml} \cite{Dutheil2008} restricted to a nonhomogeneous continuous-time K81 model. On fig. 3 (right) we present the results of these two methods on nonhomogeneous K81 data to show that both methods perform similarly and outperform \texttt{NJ}. 

Although \texttt{EMtree} may give more accurate results on more general data, it is a time consuming algorithm. Indeed, on 4-taxon trees, \texttt{EMtree} is almost 1000 times slower than \texttt{NJ} and more than 2000 times slower than \texttt{ML}. For example on 100 4-taxon alignments of 600bp,  the execution time on a  Intel(R) Core(TM) i5-4200U CPU @ 1.60GHz (only using one of the four CPU) was 
186.66s for \texttt{EMtree} on K81*, 0.06s for PAML \texttt{ML} on homogeneous continuous-time K81, 211.45 for \texttt{bppml} on nonhomogeneous K81 model, and 0.16s for \texttt{NJ} (with K81 distances).

Now we present the results that test the use of \texttt{EMtree}, \texttt{NJ}, and  (time-homogeneous continuous-time) \texttt{ML}  as methods to obtain the input for the quartet-based methods QuartetSuite and QFIT on data generated under K81* model.

For the three topologies on twelve taxa studied here (called \textit{cc, cd}, and \textit{dd} --see figure 4 and the Methods section), we give the proportion of the (one thousand) reconstructed trees whose Robinson-Foulds distance to the original topology is equal to $0, 2 , 4, 6$ or $>6$, for both QuartetSuite and QFIT, and for the three input methods under study. The results are displayed in figures 5
% \ref{fig:quartetsuite_lon600}
(QuartetSuite) and 6
% \ref{fig:clann_lon600}
(QFIT) for simulated alignments of 600 bp and in figures A1 and A2 of the Additional file for 300 bp. The figures for both alignment lengths are similar and present the same trends (slightly better for 600 bp for all methods, as expected), so we let the results on 300 bp for the additional file. In figure 5 we also present the performance of a global \texttt{NJ} and a global \texttt{ML} (estimating time-homogeneous continuous-time K81 model).

In both quartet-based methods (QuartetSuite and QFIT), the reconstruction of the \textit{cc} topology  presents the best results compared to the others, independently of the  algorithm used as input quartets (\texttt{EMtree}, \texttt{NJ}, or \texttt{ML}). The same tendency is shared by global \texttt{ML} and \texttt{NJ}. Conversely, the topology $dd$ is never correctly reconstructed for any of the methods or branch lengths. It is worth noting that $cc$ is the topology that would have the least long branch attraction and $dd$ is the one that would have the most ($cd$ is in between because half of the tree comes from $cc$ and the other half comes from $dd$). Therefore our results are consistent with the observation in \cite{Ranwez2001} that the success of a quartet based method depends on the capacity of the input method to correctly reconstruct 4-taxon trees under the long-branch attraction problem.

In both figures 5
% \ref{fig:quartetsuite_lon600}
and 6
% \ref{fig:clann_lon600}
we observe that the performance of \texttt{ML} and \texttt{EMtree} is quite similar in most cases, although \texttt{ML} never outperforms \texttt{EMtree}. A detailed look at these figures reveals that for largely  divergent trees (that is, $b=0.1$, the last bar in each plot), \texttt{EMtree} is the best quartet input method among those considered here, as its results outperform \texttt{NJ} and \texttt{ML} for both QuartetSuite and QFIT and in all tree topologies. We find the explanation of this result in the management of long-branch attraction by the different methods considered. Long branches lead to similar sequences as a result of multiple substitutions and, as \texttt{ML} estimates have been computed on a wrong substitution model, this method  is more influenced by long-branch attraction.
%as mentioned above, \texttt{EMtree} is the algorithm that deals better with this problem.
\texttt{EMtree} has also a better success than a global   \texttt{NJ} and a global \texttt{ML} on the trees $cd$ and $dd$, but on the ``easiest'' tree $cc$, a global \texttt{ML} is more accurate. For the largely divergent $cc$ tree, a global \texttt{ML} has better average performance than \texttt{EMtree} (see also below), but it never succeeds in fully reconstructing the topology (whereas \texttt{EMtree} does in about 2\% of the alignments).    

 When considering the QuartetSuite method, \texttt{NJ} is clearly the worst quartet input method for all tree topologies and branch lengths. This is probably due to the fact that  the Willson method implemented in QuartetSuite is intended for weighted quartets, whereas \texttt{NJ} quartets are given only binary weights. On the contrary, for the unweighted method QFIT, \texttt{NJ}  seems to be more accurate than \texttt{ML} and \texttt{EMtree} for topologies \textit{cc} and \textit{dd} but only for low divergence ($b=0.005$ or $b=0.015$).

 % Based on our results for the QuartetSuite algorithm, EMtree and ML perform better than NJ. However, Clann gives worse results and NJ seems work better than other methods for the \textit{cc} topology (figs \ref{fig:quartetsuite_lon300},\ref{fig:quartetsuite_lon600}, \ref{fig:clann_lon300},\ref{fig:clann_lon600}). These results could depend on the parameters chosen for Clann and QuartetSuite????????????????

We want to point out that, in general, QFIT gives  worse results than QuartetSuite (except for the \texttt{NJ} algorithm used as input in the \textit{cc} and \textit{cd} topologies with low divergence), reinforcing the idea that weighted methods are more reliable \cite{Ranwez2001}, \cite{Strimmer1997}.

In tables 1 and 2 we display the mean of the Robinson-Foulds distance of the same study on the 12-taxon trees, and its variance in parentheses (table 1%\ref{tb:mean1_quartet600} 
for QuartetSuite, and table 2
%\ref{tb:mean1_clann600} 
for QFIT). Results are presented for alignments of 600 bp and have to be interpreted as less mean, better approximation (results for 300 bp are similar and appear in the additional file, tables S1 and S2). For each tree topology and each choice of parameter $b$, the best method (according to the lowest mean) is marked in bold print.

In these tables we observe that QuartetSuite gives the lowest distance to the original tree in general and that the best results for largely divergent trees ($b=0.1$) are obtained by \texttt{EMtree} (for both QuartetSuite and QFIT and all tree topologies). Whenever the mean of Robinson-Foulds distance for (quartet-based) \texttt{ML} is lower than the mean for \texttt{EMtree}, there is no significant difference. Overall, \texttt{EMtree} is the method that outperforms the other two quartet input methods in most cases. As far as global methods are concerned, it is worth pointing out the bad performance of a global NJ on trees $cd$ and $dd$ (or $cc$ with $b=0.05$ or $0.1$), specially if one takes into account that the Robinson-Foulds distance for the (less resolved) star tree is 9.

When inspecting the variances, one sees that \texttt{EMtree} is the only quartet input method that preserves low variances in all cases (global methods are the ones presenting lower variance, though). Conversely, the variances are extremely large for \texttt{NJ} with QuartetSuite in all different scenarios, and they are also huge for (quartet input) \texttt{ML} when the trees are largely divergent ($b=0.1$). QFIT presents low variances in all cases, for all input methods, probably because the input information is less ``subject to vary'' (the input only considers tree topologies, not weights). It is worth pointing out that whenever quartet input \texttt{NJ} outperforms \texttt{ML} and \texttt{EMtree} (only for QFIT), it does so with larger variance than the other two methods.

\section*{Conclusion}

We tested the accuracy of the (likelihood-based) method \texttt{EMtree} as a method to infer 4-taxon topologies under (time nonhomogeneous) Markov models, and compared it to \texttt{NJ} and \texttt{ML} (homogeneous and nonhomogeneous). When \texttt{EMtree} and \texttt{ML} are tested on data satisfying the assumptions of both methods, they have a similar performance. Nevertheless, \texttt{EMtree} is based on time nonhomogeneous models (both local and global time heterogeneity), and hence outperforms the other methods when these assume homogeneity. 
%it outperforms \textt{NJ} and time-homogeneous \texttt{ML}, specially in the presence of long branches.
%It turns out that, when data do  not satisfy the homogeneity assumption (that is, sequences do not share a common instantaneous rate matrix), \texttt{EMtree} outperforms the other two methods specially regarding the long-branch attraction phenomenon. 
There are only few nonhomogeneous continuous-time models that could be fairly compared to general Markov processes, and
%for them there is no software publicly available. For example, we did not have at our disposal software to estimate nonhomogeneous continuous-time K81 model, where one could test the effect of local homogeneity that underlies all continuous-time models (but not Markov processes in general). 
one expects that under more complex evolutionary scenarios (such as non-stationary or not time-reversible data), the success of usual \texttt{ML} or \texttt{NJ} methods (based on these assumptions) will be poorer (as shown in \cite{Ho2004}), confirming that an EM approach based on general Markov processes, could be more recommendable. 

\texttt{EMtree} is a time-consuming algorithm, however, and the user has to decide whether it is worth performing such an analysis (we only recommend it for at most 6 taxa).

We have also assessed \texttt{EMtree}, \texttt{NJ}, and \texttt{ML} as input for the quartet-based methods QFIT and QuartetSuite. To do so, we have considered three different topologies on twelve taxa evolving under tome nonhomogeneous processes, and a wide set of branch lengths values. \texttt{EMtree} turns out to be the input method that performs best in most cases on this type of data. Regardless of the quartet-based method chosen and the tree topology, \texttt{EMtree} gives the best results for trees with large divergence among taxa ($b=0.1$). 
%, precisely where the long-branch attraction problem is more visible.

Summing up, an EM approach on Markov models provides an accurate 4-taxon tree reconstruction method suited for data not known to satisfy homogeneity and very useful as input of quartet-based methods, specially for largely divergent trees. However, the method presented here is not valid for data violating the i.i.d. hypotheses, such as data with variation across sites (Gamma-rates, invariable sites, mixtures, and others),  or dependency among sites. The method should be strongly modified in order to accommodate these generalizations, and this might be an interesting future project.

\section*{Methods}

\subsection*{EM algorithm}
We have implemented an expectation-maximization (EM) algorithm on four-taxon trees evolving under Markov processes. 
The core of the EM algorithm for phylogenetic trees is:
\begin{itemize}
 \item[-] \textit{Input:} Data $D$ (multiple alignment of $n$ sequences), unrooted trivalent tree topology $T$ with $n$ leaves (and $\mathfrak{e}:=2n-3$ edges), Markov model $\mathcal{M}$.
 \item[-] \textit{Initialization:} Root the tree at some internal node and provide tentative initial values for the root distribution $\pi$ and the substitution matrices $S_i$ ($i=1,\dots, \mathfrak{e}$), and a threshold $\varepsilon >0$.
 \item[-] \textit{Recursion:}
 \begin{itemize}
 \item \textit{E-step:} Provide complete data $cD$ (for all nodes in the tree) that maximizes the posterior probability $P_{T, \mathcal{M}}(S_i, \pi|cD)$ (unique maximum that can be computed efficiently by Felsenstein's algorithm \cite{Felsenstein2004}).
 
  \item \textit{M-step: }Compute the parameters $\hat{S_i}$,$\hat{\pi}$ that maximize the loglikelihood $l(\{S_i\}_i, \pi)=P_{T, \mathcal{M},\{S_i\}, \pi}(cD)$ for these complete data (this maximum is unique and can be computed in a closed form, see the additional file, section 2).

 \item If $l(\{\hat{S_i}\}_i,\hat{\pi})-l(\{S_i\}_i, \pi)>\varepsilon$, then set  $\pi=\hat{\pi}$, $S_i= \hat{S_i}$ ($i=1,\dots, \mathfrak{e}$),  and go back to \textit{E-step}.
\end{itemize}
\item[-] \textit{Output:} $\{\hat{S_i}\}_i,\hat{\pi}$ and $l(\{\hat{S_i}\}_i,\hat{\pi})$.
\end{itemize}

We set up the EM algorithm with $\varepsilon=10^{-3}$ and forced it to stop after 100 iterations if it had not finished.

Given four aligned nucleotide sequences \textit{s1, s2, s3,  s4},  we run the EM algorithm for the three possible trivalent trees on four taxa: $T_1=(s1, s2 | s3, s4)$, $T_2=(s1, s3 | s2, s4)$, $T_2=(s1, s4 | s2, s3).$  Our tree reconstruction method returns the tree topology for which the likelihood output by EM is higher. This method is called \texttt{EMtree} throughout the paper.

\subsection*{Models}
To test the method proposed in this paper, we mainly used the K81* model on four-leaved trees.  This is, the evolutionary process is a Markov process with uniform distribution at the root and is specified by  transition matrices of type
$$
             \left( \begin{array}{ccccc}
        a_i    &b_i & c_i    &d_i \\
            b_i   & a_i  &d_i   & c_i \\
             c_i  &  d_i  &a_i   & b_i \\
             d_i &   c_i  &b_i  &  a_i   \end{array} \right)
$$
on each branch $e_i$. In the matrix above, the rows and columns are labeled by nucleotides adenine, cytosine, guanine and thymine (in this order), so that entry $(j,k)$ stands for the conditional probability that nucleotide $j$ at the parent node of edge $e_i$ is substituted by nucleotide $k$ at the child node. When trees evolve under this model, the root becomes unidentifiable (different root locations may give rise to the same joint distribution at the leaves, \cite{AllmanRhodes_chapter4}) and, as a consequence, one can only expect to reconstruct unrooted trees. This model does not assume a constant instantaneous mutation rate matrix over the tree and, therefore, trees evolving under this model are time nonhomogeneous. The more restrictive models K80* and JC* (Markov versions of Kimura 2-parameter and Jukes-Cantor models) are obtained by imposing $b_i=d_i$ and $b_i=c_i=d_i$, respectively \cite[$\S$ 4.2]{AllmanRhodes_chapter4}. All of them are time-reversible (and hence stationary),
 and are part of the so-called group-based models (\cite{Evans1993}, \cite{szekely93}). The most general model among Markov processes is the \textit{general Markov model} GMM, which considers transition matrices and root distribution without any further restriction and is neither stationary nor time-reversible \cite{AllmanRhodes_chapter4}.

The other two methods that are confronted to \texttt{EMtree} in this paper are Neighbor-Joining (\texttt{NJ}) and a usual continuous-time maximum-likelihood (\texttt{ML}). The ML estimates for homogeneous continuous-time K81 model \cite{Kimura1981} (with a instantaneous mutation rate matrix constant over the tree) have been obtained by the free software PAML \cite{Yang1997} (\texttt{baseml} program), whereas for nonhomogeneous continuous-time K81 model we used the program bppml in the Bio++ package \cite{Dutheil2008}. Neighbor-Joining was implemented considering the K81 distance \cite{Kimura1981}. When a global \texttt{ML} was used on 12 taxa, we used PAML and set up the stepwise addition option on this software.

% @@@@ Diria que es l'arxiu baseml.ctl que conte
%
% \begin{verbatim}
%  seqfile = phy.txt
%  outfile = mlb
%  treefile = trespossib.txt
%  noisy = 0
%  verbose = 0
%  runmode = 0
%
% model = 10 [2 (AC CA GT TG) (AG GA CT TC)]
%  fix_kappa = 0
%  kappa = 2.0
%  getSE = 0
%  clock = 0
%  cleandata = 0
%  method = 0
%  nhomo = 0
% \end{verbatim}

\subsection*{Simulations}
The simulated data in this paper has been produced using the program GenNon-h of \cite{GenNonh} and Seq-Gen \cite{Rambaut1997}. GenNon-h  produces directly transition matrices  of the required branch length (for any of the Markov models described above) and therefore does not assume time-homogeneity (not globally over the tree, nor locally at the edges). 
Seq-Gen was used to generate data evolving under continuous-time K81 model (both homogeneous and globally nonhomogeneous). In order to generate nonhomogeneous data, we made the software generate various edges evolving at different instantaneous rates by recording ancestral sequences. Alternatively, the software $Hetero$ \cite{Jermiin2003} could be also used to generate (global) time nonhomogeneous continuous-time data.
%In other unpublished tests, we observed that the results of \texttt{ML} and \texttt{NJ} slightly improve if we use software that generates homogeneous data, but \texttt{EMtree} does not loose accuracy and the tendency of ML and NJ in the different setups tested here is analogous to the one explained above.

\subsubsection*{Tree space on 4-leaf trees}
In order to test the performance of the proposed method on four-taxon trees, we based our tests on the tree space proposed by Huelsenbeck \cite{Huelsenbeck1995}, so that it is possible to compare our results to those obtained there with different phylogenetic reconstruction  methods.  In this tree space, two branch length parameters $a$, $b$ on trees of four taxa are varied.  Parameter $a$ assigns the branch length to the
internal branch and two opposite peripheral branches, and parameter $b$ assigns the branch length to the
two remaining branches as in figure 2.
% \ref{fig:arbre}
Parameters $a$ and $b$ represent expected number of substitution per site and in this paper are varied from 0.01 to 0.75 in increments of 0.02.

For each pair $(a,b)$, we simulated one hundred alignments of lengths 300 and 1000 basepairs (briefly bp) under the tree topology $12|34$ (see figure 2)
% \ref{fig:arbre}
and we inferred the topology using the three methods \texttt{EMtree}, \texttt{ML}, and \texttt{NJ}. The results for each method are shown in figure 1.
% \ref{fig:comparacio}

\subsubsection*{Quartet-based methods}

In order to assess the four-taxon tree reconstruction method \texttt{EMtree} proposed above, it is important to test its performance in quartet-based reconstruction methods. To do so, we considered two of these methods: Maximum Quartet Fit (QFIT) and the method proposed by Willson in \cite{Willson1999}. QFIT choses the supertree that shares the maximum number of quartets with the source trees, whereas the other method choses the supertree that optimizes a certain criterion based on the weights assigned at the input quartets.
We use the implementation of QFIT distributed in Clann \cite{Creevey2005} and the implementation of Willson's method called QuartetSuite in \cite{QuartetSuite}. While QuartetSuite produces a tree that is expected to minimize inconsistencies with quartets, QFIT uses heuristics to search through the tree space (and we restricted this search to 100000 trees).
 As our intention was not to evaluate quartet-based methods but testing our  4-taxon method in comparison to others,  our criterion to choose these two quartet-based methods was the fact that they were freely available and that one of them allowed the use of weighted quartets.

In QuartetSuite weights are understood as -log of the frequency of a quartet, so that weights are nonnegative and zero denotes the tree with most support. Therefore we used  the opposite of the loglikelihood output by \texttt{EMtree} and \texttt{baseml} as weights of the corresponding quartets. Weights for \texttt{NJ} were set up to be 0 for the topology output by \texttt{NJ} and 999 for the other two trees.

We followed \cite{Ranwez2001} to test the performance of our method as input of quartet-based method. In that paper, the authors consider three tree topologies on 12 taxa, denoted as \textit{cc}, \textit{cd} and \textit{dd} (see figure 4),
% \ref{fig:topologies}
and fix the proportions among their branch lengths in order to compare different reconstruction methods. A parameter $b$ denoting the internal branch lengths is varied between $0.005$, $0.015$, $0.05$,  and $0.1$, which gives a maximum pairwise divergence along the tree of about $0.1$, $0.3$, $1.0$, and $2.0$ nucleotides per site, respectively. The  lengths of the simulated alignments are 300 and 600 bp.

For each of these scenarios we generated 1000 alignments using GenNon-h and the Kimura 3-parameter model as explained above. For each alignment, the tree was estimated using QFIT and QuartetSuite with the four-taxon methods \texttt{EMtree}, \texttt{ML}  and \texttt{NJ} as input. Then the Robinson-Foulds distance to the original tree (that is, the number of partitions that are present in one tree but not in the other) was computed using DendroPy symmetric\_difference function \cite{suku}. The results are shown in figures 5
% \ref{fig:quartetsuite_lon600}
, 6
% \ref{fig:clann_lon600}
, A1 and A2 (in the additional file), and in tables  1, 2, and S1, S2 (in the additional file), and explained in the Results section.

%\subsection*{Sub-sub heading for section}
%\subsubsection*{Sub-sub-sub heading for section}

\bigskip

%%%%%%%%%%%%%%%%%%%%%%%%%%%%%%%%
\section*{Competing interests}
  The authors declare that they have no competing interests.

%%%%%%%%%%%%%%%%%%%%%%%%%%%%%%%%
\section*{Author's contributions}
    The first author implemented the methods, produced results, and wrote the paper. The second author had the idea, directed the
    work, implemented methods and wrote the paper.

%%%%%%%%%%%%%%%%%%%%%%%%%%%
\section*{Acknowledgements}
  \ifthenelse{\boolean{publ}}{\small}{}
  MC acknowledges funding from Ministerio de Ciencia y Competitividad of Government of Spain, project MTM2012-38122-C03-01, and Generalitat de Catalunya, 2009 SGR 1284. We thank the editor and reviewers for highly valuable comments that lead to major improvements of the manuscript.

%%  Bmc_article.bst  will be used to                       %%
%%  create a .BBL file for submission, which includes      %%
%%  XML structured for BMC.                                %%
%%  After submission of the .TEX file,                     %%
%%  you will be prompted to submit your .BBL file.         %%

% \newpage
% {\ifthenelse{\boolean{publ}}{\footnotesize}{\small}
%  \bibliographystyle{bmc_article}  % Style BST file
%   \bibliography{emtree} }     % Bibliography file (usually '*.bib' )

%%%%%%%%%%%

\ifthenelse{\boolean{publ}}{\end{multicols}}{}

%%%%%%%%%%%%%%%%%%%%%%%%%%%%%%%%%%%
%%                               %%
%% Figures                       %%
%%                               %%
%% NB: this is for captions and  %%
%% Titles. All graphics must be  %%
%% submitted separately and NOT  %%
%% included in the Tex document  %%
%%                               %%
%%%%%%%%%%%%%%%%%%%%%%%%%%%%%%%%%%%

%%% noves

\section*{Figures}
%   \subsection*{Figure 1 - Sample figure title}
%       A short description of the figure content
%       should go here.

  \subsection*{Figure 1 %\ref{fig:comparacio}
  - Results of three different topology reconstruction methods: \texttt{NJ}, \texttt{ML} and \texttt{EMtree}\label{fig:comparacio}}
    Results of three different topology reconstruction methods (from top to bottom: \texttt{NJ}, \texttt{ML} and \texttt{EMtree}) on the tree space proposed by Huelsenbeck on K81* data (x-axis corresponds to parameter $a$ in Figure 2
%     \cite{fig:arbre}
    and y-axis to parameter $b$ in the same figure). Black cell is drawn if the topology is 100\% correctly reconstructed on the simulated data. White square denotes 0 to 33\% success, and the gray scale in between. Left figures correspond to alignments of lengths 300 bp and right to 1000 bp.

%   \subsection*{Figure \ref{fig:quartetsuite_lon300} - Results of QuartetSuite (quartet-based method) with alignment length equal to 300.}
%     Results of QuartetSuite (quartet-based method) from different topologies with length alignments equal 300. From the top to the bottom (a,b,c) correspond to different topologies: cc, cd, dd. From the left to right: Using NJ, ML, EMtree (respectively) methods to obtain quartets to give as input to QuartetSuite.
  \subsection*{Figure 2 %\ref{fig:arbre} 
  - Tree with two parameters as branch lengths.\label{fig:arbre}}
   Four-taxon tree used to simulate data in the Huelsenbeck tree space. Parameter $a$ denotes the length of the inner and two outer branches and parameter $b$ is the length of the other peripheral branches. Branch length is measured as the expected number of substitutions per site and varies between $0.01$ and $0.75$ in our simulations.

  \subsection*{Figure 3%\ref{fig:novafig3}
  - Results of three different topology reconstruction methods: \texttt{NJ}, \texttt{ML} and \texttt{EMtree}\label{fig:novafig3}}
Results of three different topology reconstruction methods from the top to the bottom \texttt{NJ}, \texttt{ML} and \texttt{EMtree} on the tree space proposed by Huelsenbeck (see fig. 2). On the left, data have been generated using a time-homogeneous continuous-time K81 model and the topology is estimated by \texttt{NJ} (with K81 distance), \texttt{ML} (estimating a homogeneous K81 model), and \texttt{EMtree} (on K81*). On the right, data have been generated using time nonhomogeneous continuous-time K81 model and the topology is estimated by \texttt{NJ} (with K81 distance), \texttt{ML} (estimating time nonhomogeneous K81 continuous-time model), and \texttt{EMtree} (on K81*). The alignments considered in this figure have length 600bp.

  \subsection*{Figure 4%\ref{fig:topologies}
  - Twelve-taxon topologies used in the paper: \textit{cc, cd ,dd}.\label{fig:topologies}}
  Twelve taxa topologies used in the paper, named according to \cite{Ranwez2001}: \textit{cc} (left), \textit{cd} (middle), \textit{dd} (right); $b$ is the branch length parameter that varies in $\{0.005, 0.015, 0.05, 0.1 \}$ along the paper.

   \subsection*{Figure 5%\ref{fig:quartetsuite_lon600}
   - Results of QuartetSuite (quartet-based method) with alignment length equal to 600.\label{fig:quartetsuite_lon600}}
   Results of QuartetSuite (quartet-based method) for different topologies and different input methods on alignments of length 600. The first three rows correspond to different methods of obtaining the quartets: \texttt{NJ} (a), \texttt{ML} (b), \texttt{EMtree} (c), and the two bottom rows correspond to global methods: (d) global \texttt{ML} (estimating homogeneous continuous-time K81 model), (e) global \texttt{NJ} (with K81 distances).  Columns correspond to the three different 12-taxon topologies simulated: \textit{cc} (left),  \textit{cd} (middle), \textit{dd} (right).

%   \subsection*{Figure \ref{fig:clann_lon300} - Results of Clann (quartet-based method) with alignment length equal to 300.}
%     Results of Clann (quartet-based method) from different topologies with length alignments equal 300. From the top to the bottom (a,b,c) correspond to different topologies: cc, cd, dd. From the left to right: Using NJ, ML, EMtree (respectively) methods to obtain quartets to give as input to Clann.

  \subsection*{Figure 6%\ref{fig:clann_lon600}
  - Results of QFIT (quartet-based method) with alignment length equal to 600.\label{fig:clann_lon600}}
    Results of QFIT (quartet-based method)  for different topologies and different input methods on alignments of length 600. Each row corresponds to a different method of obtaining the quartets: \texttt{NJ} (a), \texttt{ML} (b), \texttt{EMtree} (c).  Columns correspond to the three different 12-taxon topologies simulated: \textit{cc} (left),  \textit{cd} (middle), \textit{dd} (right).

%%%%%%%%%%%%%%%

%%%%%%%%%%%%
%%%%%%%%%%%%%%%%%%%%%%%%%%%%%%%%%%%
%%                               %%
%% Tables                        %%
%%                               %%
%%%%%%%%%%%%%%%%%%%%%%%%%%%%%%%%%%%

%% Use of \listoftables is discouraged.
%%
\section*{Tables}
%   \subsection*{Table 1 - Sample table title}
%     Here is an example of a \emph{small} table in \LaTeX\ using
%     \verb|\tabular{...}|. This is where the description of the table
%     should go. \par \mbox{}
%     \par
%     \mbox{
%       \begin{tabular}{|c|c|c|}
%         \hline \multicolumn{3}{|c|}{My Table}\\ \hline
%         A1 & B2  & C3 \\ \hline
%         A2 & ... & .. \\ \hline
%         A3 & ..  & .  \\ \hline
%       \end{tabular}
%       }

%%%%%%%%%% taules dividides en 2 cadasacuna %%%%%
% \newpage
  
  \subsection*{Table 1 - Mean of Robinson-Foulds distance with QuartetSuite, 600 bp.}
 Mean of Robinson-Foulds distance with QuartetSuite and global \texttt{NJ} and \texttt{ML}. Variance is shown in brackets and $b$ refers to the branch length parameter (see Figure 4%\ref{fig:topologies}
 ). The length is 600 bp for the 1000 sampled alignments. The inputs for QuartetSuite have been obtained from the different methods: \texttt{NJ}, \texttt{ML}, \texttt{EMtree}. For each tree topology and each choice of parameter $b$, the best results are marked in boldface.
  \par \mbox{}
    \par
    \mbox{
 \begin{tabular}{l|l|llll}
  \multicolumn{6}{l}{Mean Robinson-Foulds distance with QuartetSuite, 600 bp}\\
    \cline{1-6}
  top. & Method&  \multicolumn{4}{ c }{b}\\
  \cline{3-6}
    & & 0.005 & 0.015 & 0.05 & 0.1 \\
    \hline
    \textit{cc} & \texttt{NJ} & 7.63 (10.82)& 6.55 (18.32) & 10.70 (13.04) & 12.82 (9.57)\\
    & \texttt{ML} & {3.70} (4.91) & {2.16} (1.46) & {3.04} (1.42) & 7.37 (19.97)\\
    & \texttt{EMtree} & 3.77 (4.14) & 2.22 (2.33) & 3.20 (1.10) & {3.62} (3.27)\\
    \cline{2-6}
    & \texttt{global NJ} & 1.68 (3.42) & 1.60 (0.64) & 6.96 (10.76) & 13.16 (1.45)\\
    & \texttt{global ML} & \textbf{1.04} (3.08) & \textbf{0.0} (0.0) & \textbf{0.0} (0.0) & \textbf{3.10} (0.99)\\
    \hline\hline
    \textit{cd} & \texttt{NJ} &  8.01 (14.01) & 8.96 (11.61) & 11.16 (12.54)& 12.31 (10.48)\\
    & \texttt{ML} & {5.16} (3.56) & \textbf{3.34} (2.03) & \textbf{3.39} (1.59) & 8.12 (19.28)\\
    & \texttt{EMtree} & 5.31 (3.96)&3.71 (2.82) & 3.67 (1.61) & \textbf{4.38} (4.12)\\
    \cline{2-6}
    & \texttt{global NJ} & 5.80 (3.72) & 5.56 (0.69) & 8.44 (0.69) & 13.12 (1.31)\\
    & \texttt{global ML} & \textbf{5.04} (3.08) & 4.0 (0.0) & 4.0 (0.0) & 6.60 (0.84)\\
    \hline\hline
    \textit{dd} & \texttt{NJ} & 10.55 (9.53) & 9.45 (9.39) & 10.90 (13.95) & 11.97 (10.91)\\
    & \texttt{ML} & 7.71 (3.67) & \textbf{6.42} (1.12) & 6.08 (1.58) & 9.58 (17.09)\\
    & \texttt{EMtree} & {7.58} (5.07) & 6.87 (1.73) & \textbf{5.94} (1.44) & \textbf{6.60} (4.60)\\
    \cline{2-6}
    & \texttt{global NJ} & 9.24 (0.94) & 9.60 (0.64) & 10.4 (2.88) & 13.08 (1.31)\\
    & \texttt{global ML} & \textbf{7.48} (0.77) & 8.0 (0.0) & 8.0 (0.0) & 10.60 (0.84)\\
    \end{tabular}
     \label{tb:mean1_quartet600}
    }

  \subsection*{Table 2 - Mean of Robinson-Foulds distance with QFIT, 600 bp.}
  Mean of Robinson-Foulds distance with QFIT. Variance is shown in brackets and $b$ refers to the branch length parameter (see Figure 4%\ref{fig:topologies}
  ). The length is 600 bp for the 1000 sampled alignments. The inputs for QFIT have been obtained from the different methods: \texttt{NJ}, \texttt{ML}, \texttt{EMtree}. For each tree topology and each choice of parameter $b$, the best results are marked in boldface.
  \par \mbox{}
    \par
    \mbox{
  \begin{tabular}{l|l|llll}
  \multicolumn{6}{l}{Mean Robinson-Foulds distance with QFIT, 600 bp}\\
    \cline{1-6}
  top. & Method&  \multicolumn{4}{ c }{b}\\
  \cline{3-6}
    & & 0.005 & 0.015 & 0.05 & 0.1 \\
    \hline
    \textit{cc} & \texttt{NJ} & \textbf{1.05} (2.22) & \textbf{0.67} (1.31) & 8.12 (8.44) & 12.97 (2.55)\\
    & \texttt{ML} & 4.68 (0.90) & 4.36 (0.98) & 4.37 (0.80) & 5.31 (2.03)\\
    & \texttt{EMtree} & 4.58 (0.98) & 4.21 (0.38) & \textbf{4.25} (0.44) & \textbf{4.66} (1.09)\\
    \hline
    \textit{cd} & \texttt{NJ} & \textbf{4.60} (1.41) & \textbf{4.47} (1.11) & 9.53 (5.68) & 11.56 (2.46)\\
    & \texttt{ML} & 6.43 (0.88) & 6.21 (0.37) & 6.28 (0.48) & 6.92 (1.18)\\
    & \texttt{EMtree} & 6.41 (0.98) & 6.11 (0.21) & \textbf{6.20} (0.36) & \textbf{6.51} (0.76)\\
    \hline
    \textit{dd} & \texttt{NJ} & \textbf{7.93} (1.11) & 8.41 (1.16) & 10.23 (2.51) & 9.96 (2.89)\\
    & \texttt{ML} & 8.17 (0.74) & 8.15 (0.27) & 8.11 (0.20) & 8.75 (1.14)\\
    & \texttt{EMtree} & 8.14 (0.63) & \textbf{8.05} (0.11) & \textbf{8.06} (0.11) & \textbf{8.44} (0.68)\\
  \end{tabular}
   \label{tb:mean1_clann600}
  }

%%%%%%%%%%%%%%%%%%%%%%%%%% FIGURES %%%%%%%%%%%%%%%%%%%%%%%%%%%%%%%%%%%%%%%%%%%%%%%%%%%

%%%%% esther
\begin{figure}[H]
  \begin{center}
  \includegraphics[scale=0.7]{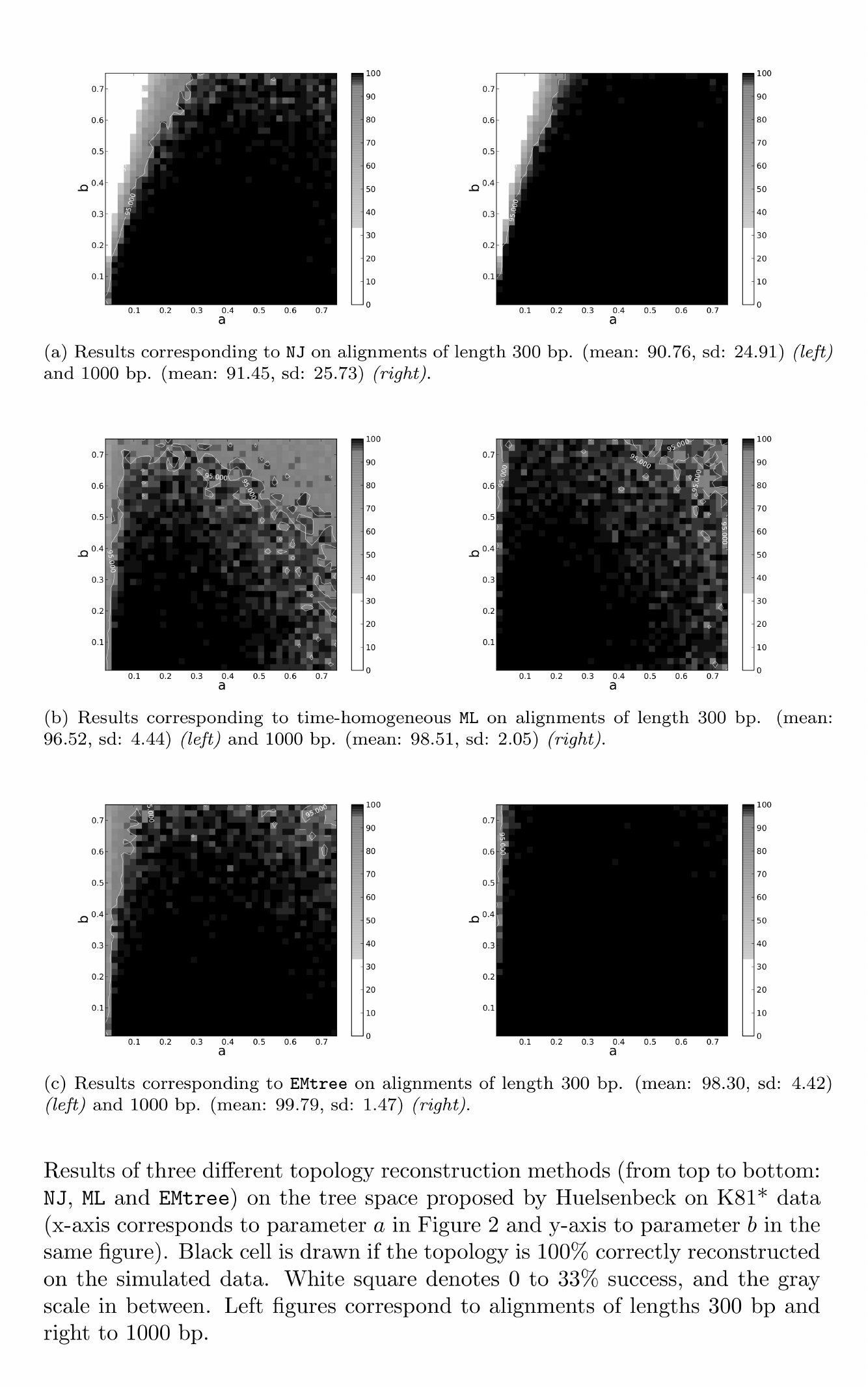}
  \end{center}
  \caption{}
%    \caption{Results of three different topology reconstruction methods (from top to bottom: \texttt{NJ}, \texttt{ML} and \texttt{EMtree}) on the tree space proposed by Huelsenbeck on K81* data (x-axis corresponds to parameter $a$ in Figure \ref{fig:arbre} and y-axis to parameter $b$ in the same figure). Black cell is drawn if the topology is 100\% correctly reconstructed on the simulated data. White square denotes 0 to 33\% success, and the gray scale in between. Left figures correspond to alignments of lengths 300 bp and right to 1000 bp.}
%   \label{fig:comparacio}
\end{figure}

\begin{figure}
  \begin{center}
  \includegraphics[width=0.6\textwidth]{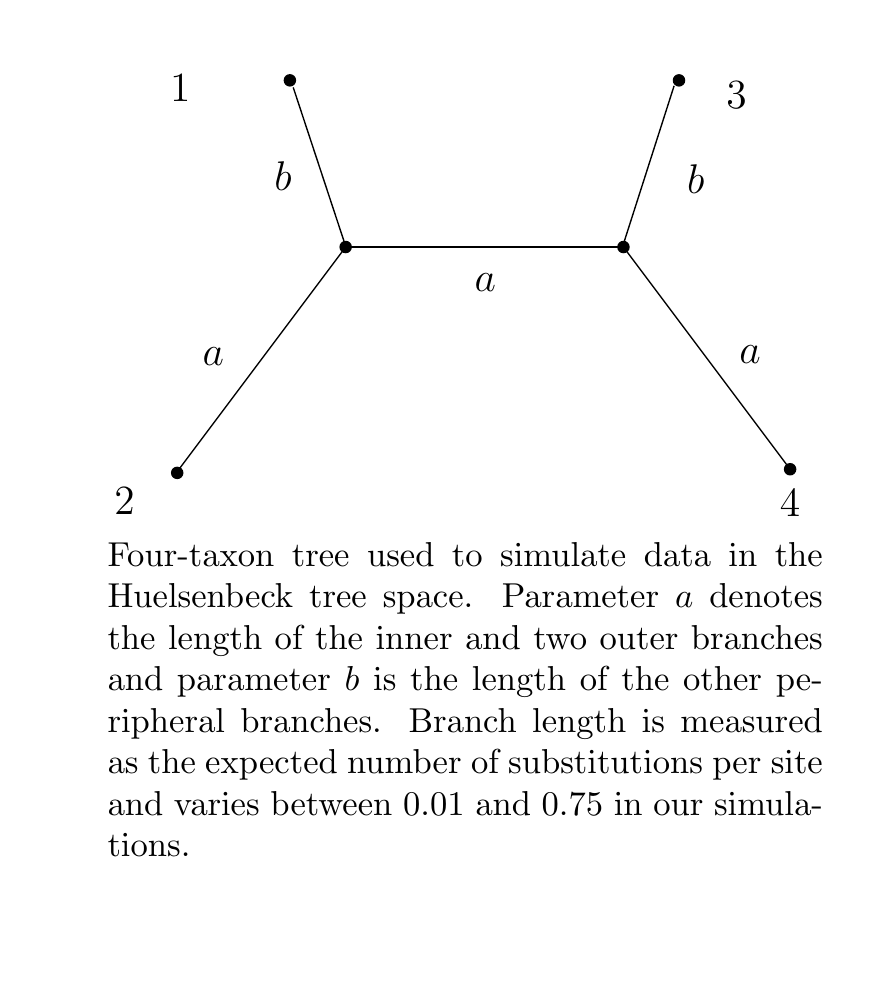}
  \end{center}
  \caption{}
%   \caption{Four-taxon tree used to simulate data in the Huelsenbeck tree space. Parameter $a$ denotes the length of the inner and two outer branches and parameter $b$ is the length of the other peripheral branches. Branch length is measured as the expected number of substitutions per site and varies between $0.01$ and $0.75$ in our simulations.}
%    \label{fig:arbre}
\end{figure}

\begin{figure}
 \begin{center}
   \includegraphics[width=0.6\textwidth]{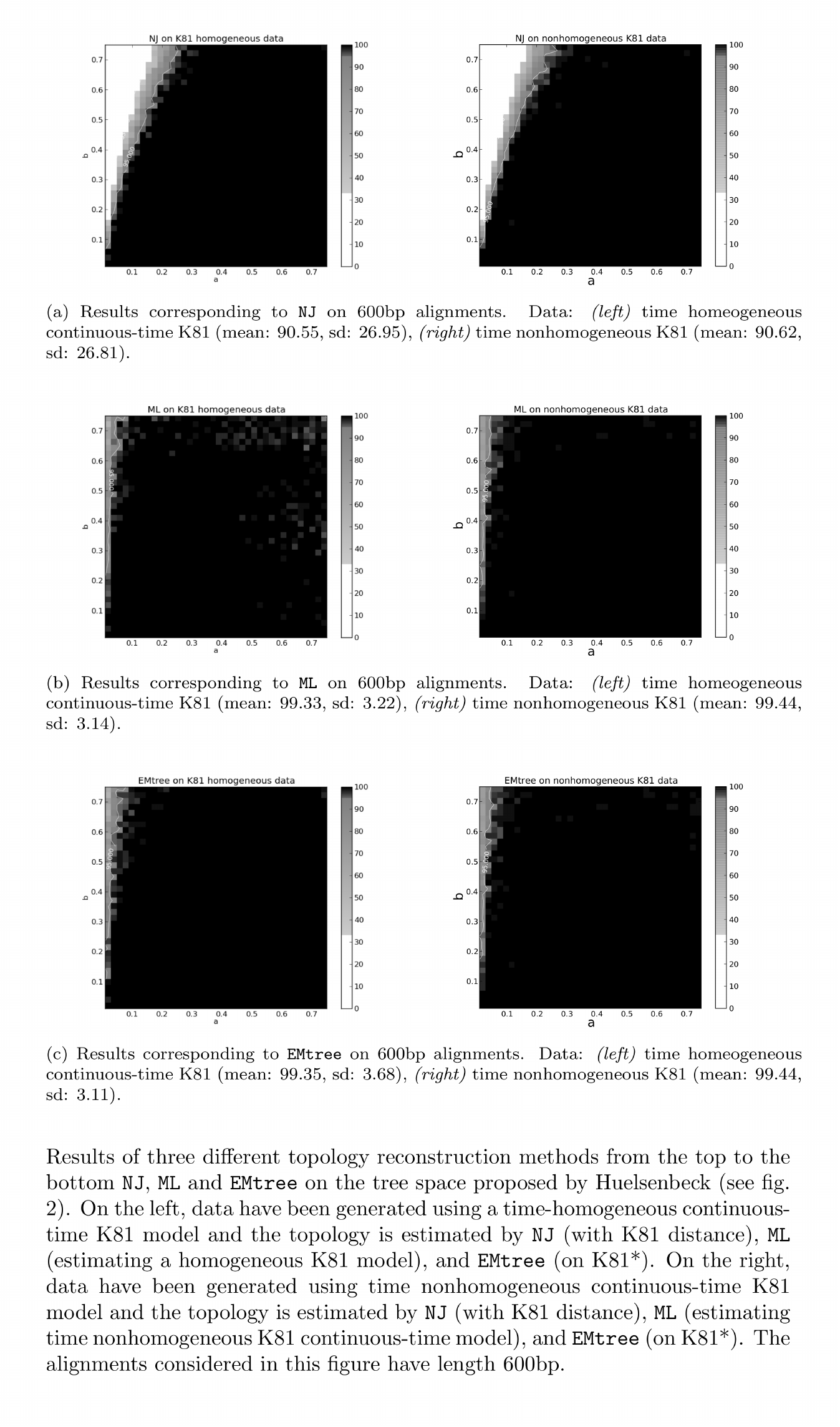}
 \end{center}
 \caption{}
%  \caption{Results of three different topology reconstruction methods from the top to the bottom \texttt{NJ}, \texttt{ML} and \texttt{EMtree} on the tree space proposed by Huelsenbeck (see fig. 2). On the left, data have been generated using a time-homogeneous continuous-time K81 model and the topology is estimated by \texttt{NJ} (with K81 distance), \texttt{ML} (estimating a homogeneous K81 model), and \texttt{EMtree} (on K81*). On the right, data have been generated using time nonhomogeneous continuous-time K81 model and the topology is estimated by \texttt{NJ} (with K81 distance), \texttt{ML} (estimating time nonhomogeneous K81 continuous-time model), and \texttt{EMtree} (on K81*). The alignments considered in this figure have length 600bp.}
% \label{fig:novafig3}
\end{figure}

\begin{figure}
 \begin{center}
   \includegraphics[width=0.7\textwidth]{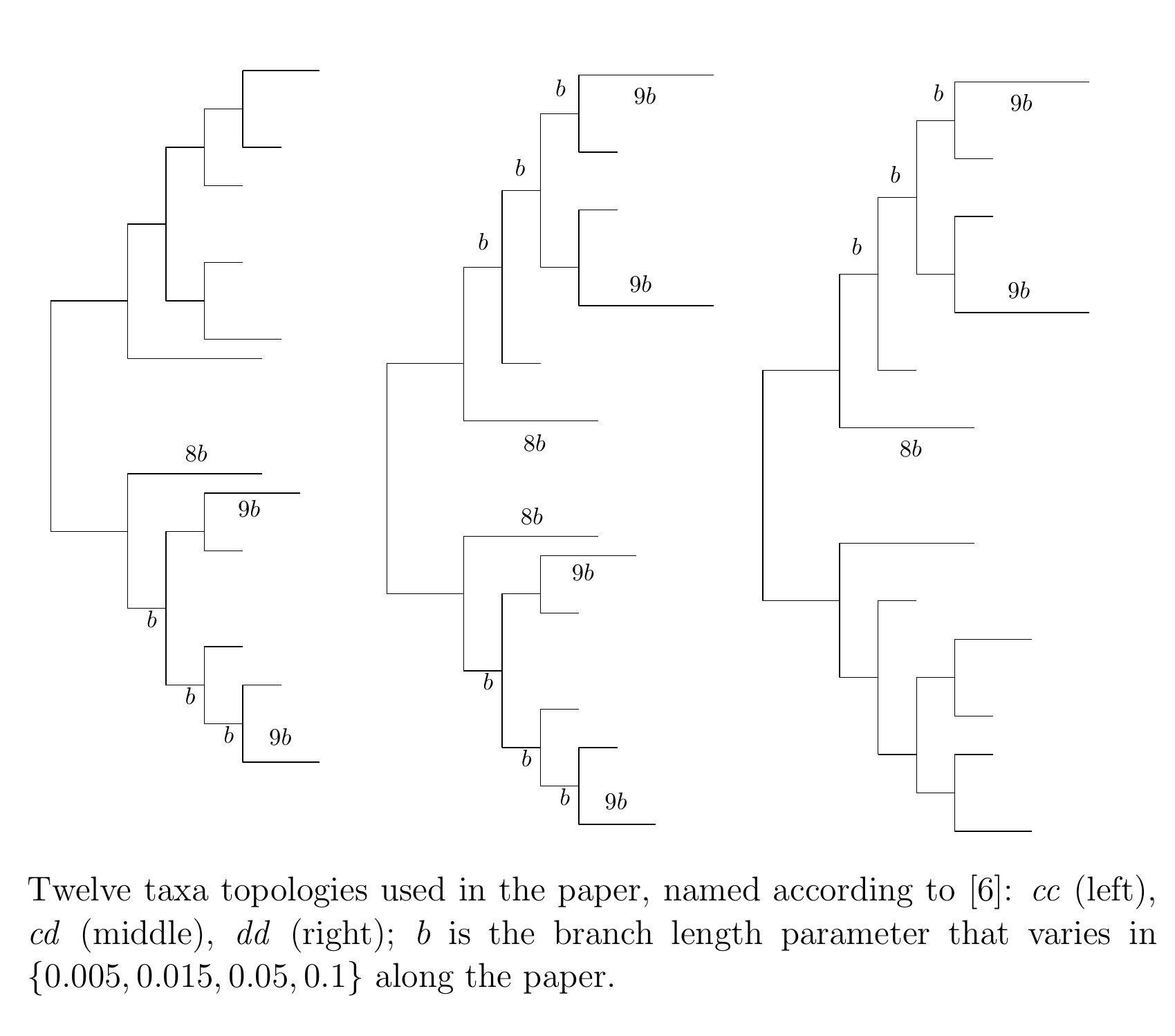}
 \end{center}
 \caption{}
%  \label{fig:topologies}
%   \caption{Twelve taxa topologies used in the paper, named according to \cite{Ranwez2001}: \textit{cc} (left), \textit{cd} (middle), \textit{dd} (right); $b$ is the branch length parameter that varies in $\{0.005, 0.015, 0.05, 0.1 \}$ along the paper.}
\end{figure}

\begin{figure}
 \begin{center}
   \includegraphics[width=0.8\textwidth]{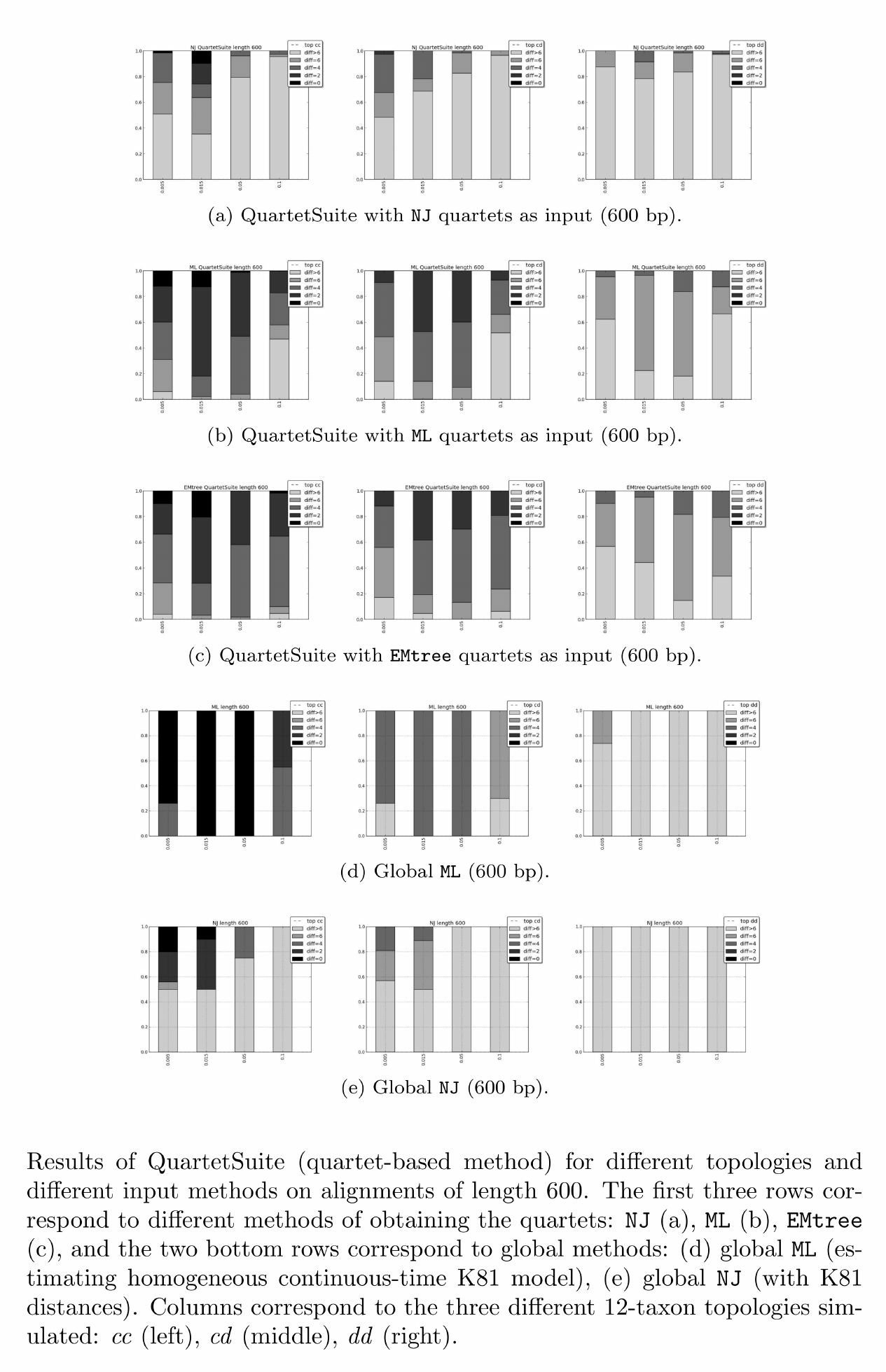}
 \end{center}
 \caption{}
%    \caption{Results of QuartetSuite (quartet-based method) for different topologies and different input methods on alignments of length 600. The first three rows correspond to different methods of obtaining the quartets: \texttt{NJ} (a), \texttt{ML} (b), \texttt{EMtree} (c), and the two bottom rows correspond to global methods: (d) global \texttt{ML} (estimating homogeneous continuous-time K81 model), (e) global \texttt{NJ} (with K81 distances).  Columns correspond to the three different 12-taxon topologies simulated: \textit{cc} (left),  \textit{cd} (middle), \textit{dd} (right).}
%   \label{fig:quartetsuite_lon600}
\end{figure}

\begin{figure}
 \begin{center}
   \includegraphics[width=0.8\textwidth]{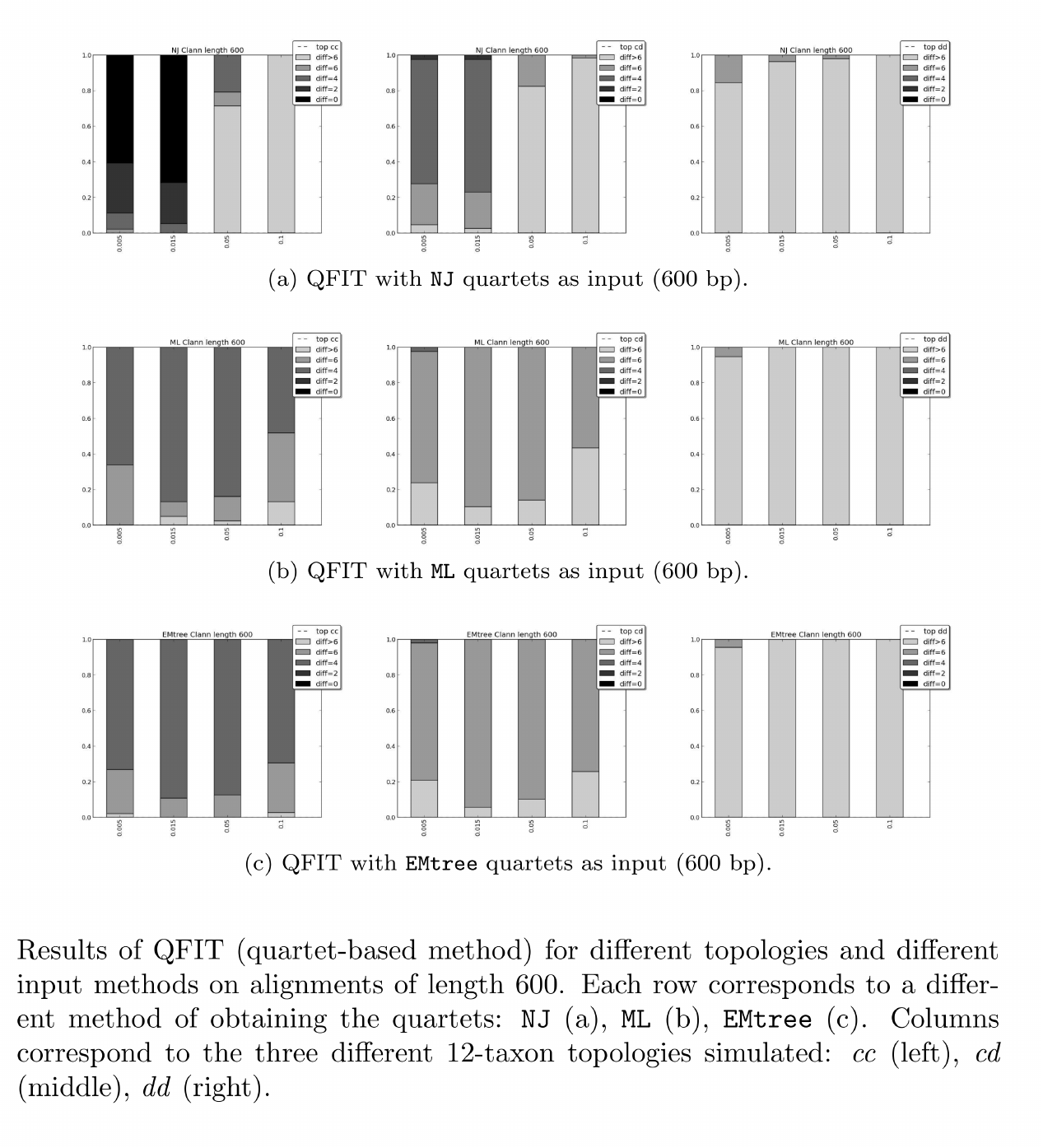}
 \end{center}
 \caption{}
%     \caption{Results of QFIT (quartet-based method)  for different topologies and different input methods on alignments of length 600. Each row corresponds to a different method of obtaining the quartets: \texttt{NJ} (a), \texttt{ML} (b), \texttt{EMtree} (c).  Columns correspond to the three different 12-taxon topologies simulated: \textit{cc} (left),  \textit{cd} (middle), \textit{dd} (right).}
% \label{fig:clann_lon600}
\end{figure}
%%%%%%%%%%%%%%%%%%%%%%%%%%%%%%%%%%%
%%                               %%
%% Additional Files              %%
%%                               %%
%%%%%%%%%%%%%%%%%%%%%%%%%%%%%%%%%%%
\section*{Additional Files}
  \subsection*{Additional file 1 --- Sample additional file title}
    Additional file is a pdf file containing figures A1, A2, and tables S1, S2 in section 1 and the explicit computation of the $M-step$ for K81* model in section 2.

\end{bmcformat}

% \documentclass[12pt]{amsart}
% %\usepackage{amssymb,latexsym,amsmath,amscd,amsthm}
% %\usepackage[latin1]{inputenc}
% %\usepackage[active]{srcltx}
% \usepackage{subfig}% per fer subfloat a les figures (esther)
% \usepackage{graphicx}% per incloure els grafics
% 
% \setlength{\textheight}{48pc}%46
% %\setlength{\textwidth}{36pc}
% \setlength{\parindent}{.4 in} \setlength{\textwidth}{145mm}
% \setlength{\topmargin} {0 in} \setlength{\evensidemargin}{1cm}
% \setlength{\oddsidemargin}{1cm} \setlength{\footskip}{.3 in}
% \setlength{\headheight}{.8 in}
% %\setlength{\textheight}{8.8 in}
% \setlength{\parskip}{.1 in}
% 
% \baselineskip=16 pt
% %%%%%%%%%%%%%%%%%%%%%%%%%%%%%%%%%%%%%%%%%%%%%%%%%%%%%%%%%%%%%%%%%%%
% \renewcommand{\thefigure}{A\arabic{figure}} 
% \renewcommand{\thetable}{S\arabic{table}} 
% %%%%%%%%%%%%%%%%%%%%%%%%%%%%%%%%%%%%%%%%%%%%5

\clearpage
\section*{Supplementary File}
%\author{Marta Casanellas}
%\address{Departament de Matemàtica Aplicada I. ETSEIB. Universitat Polit\`ecnica de Catalunya. Avinguda Diagonal 647. 08028 Barcelona. Spain.}
%\email{marta.casanellas@upc.edu}
%\author{Esther Iba\~{n}ez}
%\address{Centre de Recerca Matem\'atica, Campus de Bellaterra, Edifici C - 08193 Bellaterra (Barcelona), Spain.}

%\maketitle
\vspace{3cm}
\begin{figure}[h]
 \begin{center}
  \includegraphics{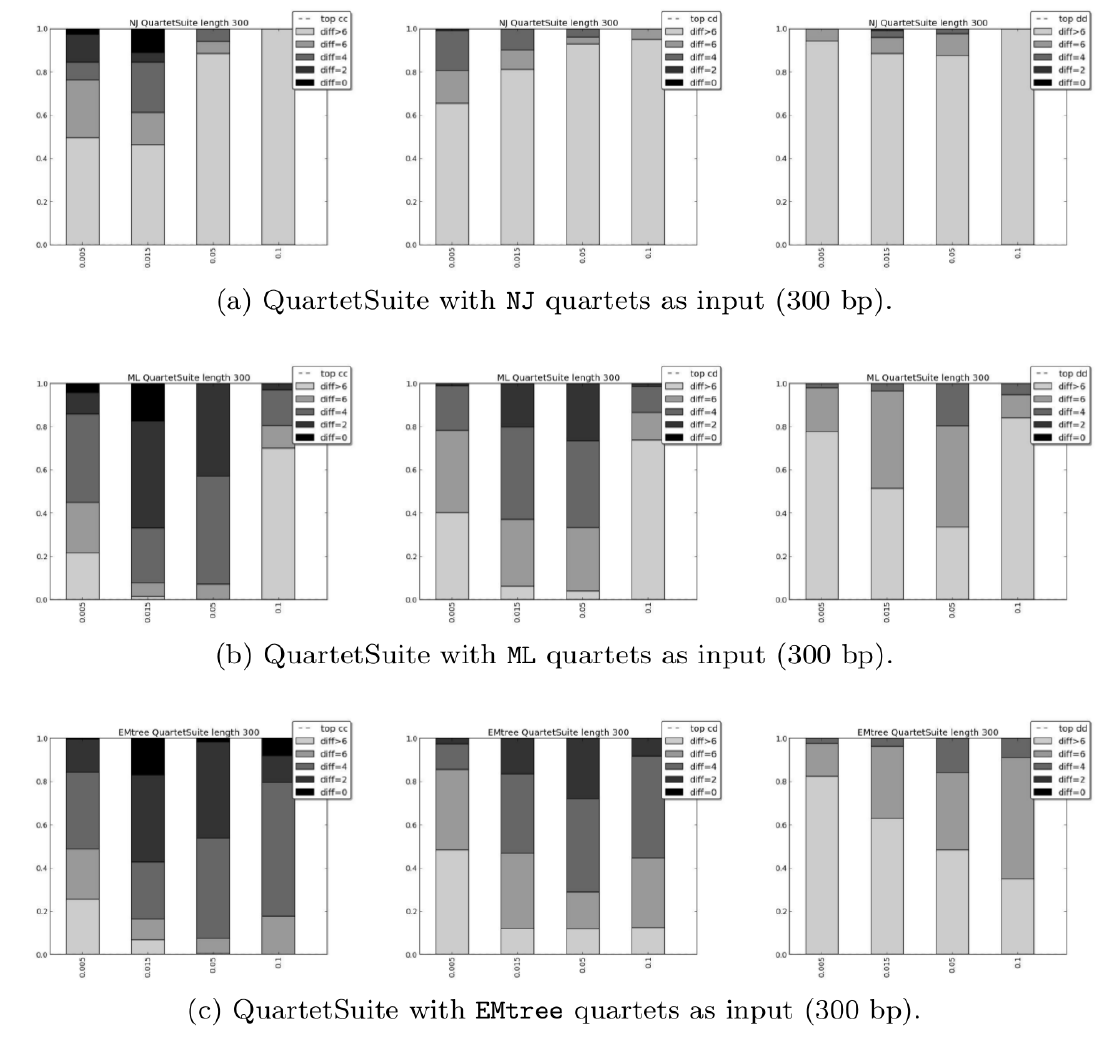}
 \end{center}
  \caption{Results of QuartetSuite for different topologies and different input methods on alignments of length 300. Each row corresponds to a different method of obtaining the quartets: \texttt{NJ} (A), \texttt{ML} (B), \texttt{EMtree} (C).  Columns correspond to the three different 12-taxon topologies simulated: \textit{cc} (left),  \textit{cd} (middle), \textit{dd} (right).}
  \label{fig:quartetsuite_lon300}
\end{figure}
% 
% \begin{figure}
%  \begin{center}
%  \subfloat[\footnotesize{QuartetSuite with \texttt{NJ} quartets as input  (300 bp).}]{
%   \includegraphics[width=0.3\textwidth]{FigureA1a1.pdf}
%   \includegraphics[width=0.3\textwidth]{FigureA1a2.pdf}
%   \includegraphics[width=0.3\textwidth]{FigureA1a3.pdf} }
% 
%  \subfloat[\footnotesize{QuartetSuite with \texttt{ML} quartets as input (300 bp).}]{
%   \includegraphics[width=0.3\textwidth]{FigureA1b1.pdf}
%   \includegraphics[width=0.3\textwidth]{FigureA1b2.pdf}
%   \includegraphics[width=0.3\textwidth]{FigureA1b3.pdf} }
%   
%   \subfloat[\footnotesize{QuartetSuite with \texttt{EMtree} quartets as input (300 bp).}]{
%   \includegraphics[width=0.3\textwidth]{FigureA1c1.pdf}
%   \includegraphics[width=0.3\textwidth]{FigureA1c2.pdf}
%   \includegraphics[width=0.3\textwidth]{FigureA1c3.pdf} }
%   \end{center}
%   \caption{Results of QuartetSuite for different topologies and different input methods on alignments of length 300. Each row corresponds to a different method of obtaining the quartets: \texttt{NJ} (A), \texttt{ML} (B), \texttt{EMtree} (C).  Columns correspond to the three different 12-taxon topologies simulated: \textit{cc} (left),  \textit{cd} (middle), \textit{dd} (right).}
%   \label{fig:quartetsuite_lon300}
% \end{figure}

\newpage

\begin{figure}[h]
 \begin{center}
  \includegraphics{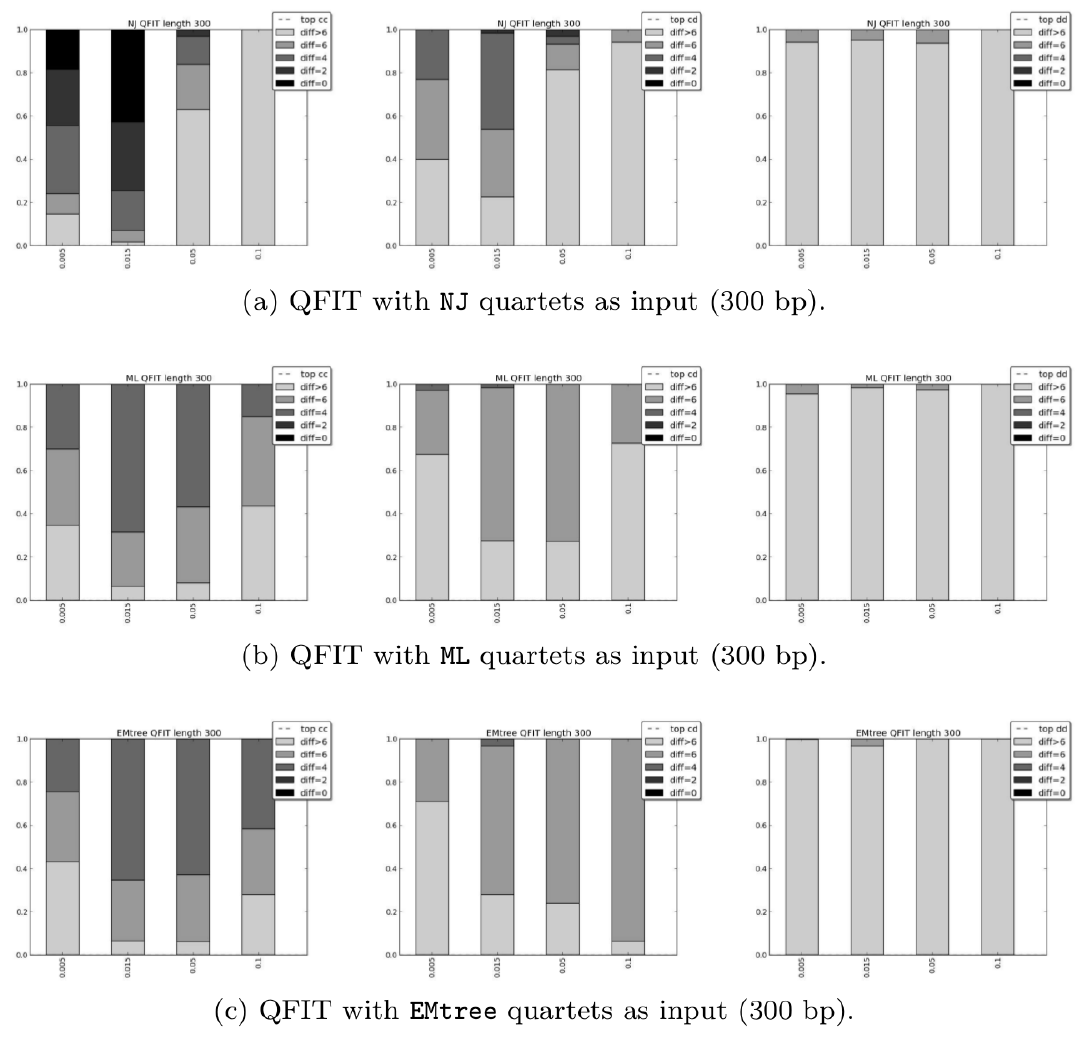}
 \end{center}
  \caption{Results of QFIT for different topologies and different input methods on alignments of length 300. Each row corresponds to a different method of obtaining the quartets: \texttt{NJ} (A), \texttt{ML} (B), \texttt{EMtree} (C).  Columns correspond to the three different 12-taxon topologies simulated: \textit{cc} (left),  \textit{cd} (middle), \textit{dd} (right).}
  \label{fig:clann_lon300}
\end{figure}

% 
%  \begin{figure}[t]
%  \begin{center}
%  \subfloat[\footnotesize{QFIT with \texttt{NJ} quartets as input  (300 bp).}]{
%   \includegraphics[width=0.3\textwidth]{FigureA2a1.pdf}
%   \includegraphics[width=0.3\textwidth]{FigureA2a2.pdf}
%   \includegraphics[width=0.3\textwidth]{FigureA2a3.pdf} }
% 
%  \subfloat[\footnotesize{QFIT with \texttt{ML} quartets as input (300 bp).}]{
%   \includegraphics[width=0.3\textwidth]{FigureA2b1.pdf}
%   \includegraphics[width=0.3\textwidth]{FigureA2b2.pdf}
%   \includegraphics[width=0.3\textwidth]{FigureA2b3.pdf} }
%   
%   \subfloat[\footnotesize{QFIT with \texttt{EMtree} quartets as input (300 bp).}]{
%   \includegraphics[width=0.3\textwidth]{FigureA2c1.pdf}
%   \includegraphics[width=0.3\textwidth]{FigureA2c2.pdf}
%   \includegraphics[width=0.3\textwidth]{FigureA2c3.pdf} }
%   \end{center}
%   \caption{Results of QFIT for different topologies and different input methods on alignments of length 300. Each row corresponds to a different method of obtaining the quartets: \texttt{NJ} (A), \texttt{ML} (B), \texttt{EMtree} (C).  Columns correspond to the three different 12-taxon topologies simulated: \textit{cc} (left),  \textit{cd} (middle), \textit{dd} (right).}
%   \label{fig:clann_lon300}
% \end{figure}

%  \subsection*{Table 1 - Mean of Robinson-Foulds distance with QuartetSuite, 300 bp.}
\newpage

\begin{table} 
\caption{
Mean of Robinson-Foulds distance with QuartetSuite. Variance is shown in brackets and $b$ refers to the branch length parameter. The length is 300 bp for the 1000 sampled alignments. The inputs for QuartetSuite have been obtained from the different methods: \texttt{NJ}, \texttt{ML}, \texttt{EMtree}. For each tree topology and each choice of parameter $b$, the best results are marked in boldface.}
\label{tb:mean1_quartet300}
\begin{center}
  \par \mbox{}
    \par
    \mbox{
 \begin{tabular}{l|l|llll}
  \multicolumn{6}{c}{Mean Robinson-Foulds distance with QuartetSuite, 300 bp}\\
    \cline{1-6}
  top. & Method&  \multicolumn{4}{ c }{b}\\
  \cline{3-6}
    & & 0.005 & 0.015 & 0.05 & 0.1 \\
    \hline
    \textit{cc} & \texttt{NJ} &  7.25 (13.02) & 6.90 (18.93) & 12.43 (16.08) & 13.69 (9.33) \\
    & \texttt{ML} & \textbf{5.39} (8.40) & \textbf{2.50} (2.99) & 3.28 (1.48) & 9.61 (18.33)\\
    & \texttt{EMtree} & 5.54 (7.73) & 3.01 (5.07) & \textbf{3.23} (2.09)& \textbf{3.79} (2.44)\\
    \hline
    \textit{cd} & \texttt{NJ} & 9.15 (15.56) & 9.94 (9.25) & 11.10 (8.58) & 13.31 (12.15) \\
    & \texttt{ML} & \textbf{7.14} (7.60) & \textbf{4.46} (2.83) & \textbf{4.22} (2.83) & 10.14 (16.82) \\
    & \texttt{EMtree} & 7.46 (7.58) & 4.88 (3.57) & 4.34 (4.48) & \textbf{4.98} (2.66) \\
    \hline
    \textit{dd} & \texttt{NJ} & 10.82 (8.43) & 11.36 (15.65) & 11.09 (12.04) & 13.01 (8.94) \\
    & \texttt{ML} & \textbf{9.14} (6.05) & \textbf{7.25} (2.51) & \textbf{6.35} (2.42) & 11.17 (14.78) \\
    & \texttt{EMtree} & 9.54 (6.95) & 7.53 (2.66)& 6.76 (2.79) & \textbf{6.70} (2.35) \\
    \end{tabular}
    
      }
      \end{center}
\end{table}

\begin{table} 
\caption{
  Mean of Robinson-Foulds distance with QFIT. Variance is shown in brackets and $b$ refers to the branch length parameter. The length is 300 bp for the 1000 sampled alignments. The inputs for QFIT have been obtained from the different methods: \texttt{NJ}, \texttt{ML}, \texttt{EMtree}. For each tree topology and each choice of parameter $b$, the best results are marked in boldface.}
   \label{tb:mean1_clann300}
   \begin{center}
  \par \mbox{}
    \par
    \mbox{
  \begin{tabular}{l|l|llll}
  \multicolumn{6}{l}{Mean Robinson-Foulds distance with QFIT, 300 bp}\\
    \cline{1-6}
  top. & Method&  \multicolumn{4}{ c }{b}\\
  \cline{3-6}
    & & 0.005 & 0.015 & 0.05 & 0.1 \\
    \hline
    \textit{cc} & \texttt{NJ}  & \textbf{3.64} (8.04) & \textbf{1.82} (3.82) & 8.10 (8.59) & 12.73 (4.02)\\
    & \texttt{ML}  &  6.67 (6.20) & 4.76 (1.46) & 5.02 (1.64) & 7.14 (5.28)\\
    & \texttt{EMtree}  &  7.05 (6.54) & 4.82 (1.48) & \textbf{4.86} (1.46) & \textbf{5.85} (3.47)\\
    \hline
    \textit{cd} & \texttt{NJ}  & \textbf{6.84} (5.70) & \textbf{5.60} (3.42) & 8.99 (6.59) & 11.86 (4.61)\\
    & \texttt{ML}  &  8.23 (4.67) & 6.53 (0.96) & 6.55 (0.79) & 8.36 (4.39) \\
    & \texttt{EMtree}  & 8.54 (4.57) & 6.52 (1.15) & \textbf{6.48} (0.73) & \textbf{7.55} (3.01)\\
    \hline
    \textit{dd} & \texttt{NJ}  &\textbf{9.59} (3.96) & 9.26 (2.69) & 10.54 (4.82) & 11.27 (3.55)\\
    & \texttt{ML}  & 9.80 (3.55) & 8.37 (0.79) & 8.43 (0.89) & 9.67 (2.79)\\
    & \texttt{EMtree}  & 9.93 (3.29) & \textbf{8.36} (0.98) & \textbf{8.33} (0.55) & \textbf{9.05} (1.97)\\
  \end{tabular}
 
  }
\end{center}
  \end{table}

\end{document}